# SPEED OF SOUND IN LIQUIDS AT HIGH PRESSURE

[1]P. Kiełczyński[a], M. Szalewski[a], S. Piekarski[b]

[a]*Section of Acoustoelectronics,*
[b]*Section of Analytical Mechanics and Field Theory*
*Department of Theory of Continuous Media,*
*Institute of Fundamental Technological Research,*
*Polish Academy of Sciences, ul. Pawińskiego 5B,*
*02-106 Warsaw, Poland.*

1) e-mail: pkielczy@ippt.gov.pl

**Abstract**

In this paper, a new general formula for the sound speed in adiabatic conditions ( $S = const$ ) has been established. The sound speed depends on the mass density $\rho(p,T)$ and the internal energy per unit mass $E(p,T)$, both expressed as functions of the pressure $p$ and the temperature $T$. This formula has been compared with experimental data on the example of triolein over the pressure range up to 450 MPa. For experimental data, phenomenological approximate formulas have been proposed. Those formulas have two versions, depending on the 2 and 3 parameters. Both versions have been developed with the help of the new expression (Eq.8) for the sound speed. The explicit form of both approximate curves can be regarded as the result of purely phenomenological modeling. However, in this paper, these new analytical expressions have been obtained by applying the heuristic procedure described in Appendix.

## I. INTRODUCTION

The ultrasonic velocity measurement has been extensively used to study the physico-chemical behavior of liquids[1-4]. In recent years ultrasonic methods have been used increasingly for determining equations of state and thermodynamic properties of liquids at elevated pressures[5-8]. Apart from the high precision attainable with ultrasonic measurements, these methods also have the advantage of simplicity and rapidity of the experimental procedure.

According to the traditional approach in acoustics, square of the velocity of sound in liquids in adiabatic conditions can be expressed by the derivative of pressure $p$ with respect to the mass density $\rho$, wherein the derivative is calculated for a constant entropy $S$: $c^2 = \frac{\partial p(\rho,S)}{\partial \rho}\Big|_{S=const}$. However, in acoustical practice we are not dealing with variables $(\rho, S)$, but usually these are the variables $(\rho, T)$ or $(p, T)$, where $T$ is temperature. Therefore, the interesting question is whether the formula (Eq.1) can be converted to a formula in the variables $(\rho, T)$ or $(p, T)$. As for the variables $(\rho, T)$ the relevant derivation was established in 2007 in Ref[9]. Whereas, in terms of variables $(p, T)$, the appropriate derivation is presented in this work.

In this work we developed a new general relation (Eq.8) that describes acoustic wave velocity in liquids in adiabatic conditions, assuming a constant entropy $S = const$, depending on the mass density $\rho(p,T)$ and internal energy per unit mass $E(p,T)$ that are expressed by the pressure $p$ and temperature $T$. To the best of our knowledge, the formula for the square of the velocity of sound in adiabatic conditions expressed in the variables $(p, T)$, is absent in the existing literature.

Mathematical modeling of the speed of sound in liquids in function of pressure is important and can simplify the determination of thermodynamic properties of liquids[10-12]. In existing so far models, the speed of sound is usually approximated using the polynomial expressions. Search of simple mathematical models for speed of sound as a function of pressure, that have potential practical applications is still an open task[12].

In this paper, the authors propose a model of the liquid in which the dependence of sound velocity on pressure is expressed through analytical expressions more general than polynomial expressions. The proposed model approximates the experimental curves more accurately with fewer parameters than corresponding polynomial approximations.

The aim of this study was developing (based on the established formula Eq.8) approximate formulas that use a small number of parameters and accurately describe the experimental results of measurements of acoustic wave velocity in a liquid as a function of pressure. Those formulas have two versions. The first version exploits the fact that the mass density and the internal energy density per unit mass can be expressed in terms of the Gibbs function. Consequently, after inserting Gibbs function depending on a small number of parameters into the formula for the sound speed one obtains the expression for sound speed that depends on a small number of parameters and the only question is whether obtained expressions give a good fit to experimental data. The second possibility is to search for individual corrections for $\rho(p,T)$ and $E(p,T)$ independently and subsequently to insert them directly into Eq.8.

In this work we used the second method, because as it results from our research, this approach provides better and more accurate results than that obtained from the first method. Different models of liquids were compared, of which particularly good fit to experimental data give two-parameter and three-parameter models. We have achieved approximations of experimental data by means of 2 and 3-parameter formulas (Eq.9 and Eq.10), which give a better fit with fewer parameters than the previously used approximate polynomial formulas[10-12]. These new analytical expressions (2 and 3-parameter curves) can be helpful in the procedure, which uses the measured values of sound speed to determine the equation of state and hence the other thermodynamic parameters.

The resulting final approximate 2 and 3-parameter formulas on the square of the velocity of sound $c^2$ can be treated as phenomenological expressions. They may have great importance for the foundations of acoustics for modeling properties of liquids exposed to high pressures. A similar approach to analyze the behavior of liquids under high pressure was reported in the paper of Holton[2]. He mentioned that analysis of the properties of liquids can be carried out in two ways, i.e., 1) using a strictly deductive methodology, or 2) applying a phenomenological approach. In the scientific research, one selects the method, which faster and easier leads to the desired results.

## II. THERMODYNAMICS OF SOUND PROPAGATION IN LIQUIDS UNDER HIGH PRESSURE

The general expression for square of sound speed for adiabatic conditions has been derived for a liquid medium with the mass density $\rho$ and the temperature $T$ taken as the independent variables[9]. However, after minor modifications, a similar formula can be derived for a fluid with the temperature $T$ and pressure $p$ taken as the independent variables. It is known that a square of the sound speed $c^2$ in adiabatic conditions is defined as a derivative of the pressure $p$ with respect to the mass density $\rho$ for the constant entropy $S$:

$$c^2 = \frac{\partial p(\rho,S)}{\partial \rho} = \frac{1}{\frac{\partial \rho(p,S)}{\partial p}} \qquad (1)$$

Our aim in this paper was to write the Eq.1 in variables $p$ and $T$. Therefore, it is convenient to write the entropy $S$ and mass density $\rho$ as the functions of $p$ and $T$, i.e., $S = S(p,T)$ and $\rho = \rho(p,T)$. To compute the sound speed for adiabatic conditions one has to consider the process that takes place for the constant entropy:

$$S(p,T) = const \qquad (2)$$

This process is isentropic and occurs on the surface of constant entropy along a curve that can be parameterized in terms of the pressure $p \to (p, T(p))$ and finally Eq.2 can be differentiated to give:

$$0 = \frac{d}{dp} S(p,T(p)) = \frac{\partial S(p,T)}{\partial p} + \frac{\partial S(p,T)}{\partial T} \frac{\partial T(p)}{\partial p} \qquad (3)$$

In turn, from Eq.3 it is possible to determine $\frac{\partial T(p)}{\partial p}$ in terms of the derivatives of entropy with respect to $p$ and $T$:

$$\frac{\partial T(p)}{\partial p} = -\frac{\frac{\partial S(p,T)}{\partial p}}{\frac{\partial S(p,T)}{\partial T}} \qquad (4)$$

Using Eq.4, the derivative in Eq.1 can be computed and presented in the form:

$$\frac{1}{c^2} = \frac{\partial \rho(p,S)}{\partial p} = \frac{\partial \rho(p,T)}{\partial p} + \frac{\partial \rho(p,T)}{\partial T}\left[-\frac{\dfrac{\partial S(p,T)}{\partial p}}{\dfrac{\partial S(p,T)}{\partial T}}\right] \qquad (5)$$

In the variables ($p$, $T$) the Gibbs identity takes the following explicit form:

$$\frac{\partial S(p,T)}{\partial p} = \frac{1}{T}\frac{\partial E(p,T)}{\partial p} - \frac{1}{T}\frac{p}{\rho^2(p,T)}\frac{\partial \rho(p,T)}{\partial p} \qquad (6)$$

$$\frac{\partial S(p,T)}{\partial T} = \frac{1}{T}\frac{\partial E(p,T)}{\partial T} - \frac{1}{T}\frac{p}{\rho^2(p,T)}\frac{\partial \rho(p,T)}{\partial T} \qquad (7)$$

Making use of the Eqs.6 and 7, it is now possible to write the expression for the sound speed in adiabatic conditions in the form:

$$\frac{1}{c^2} = \frac{\partial \rho(p,T)}{\partial p} + \frac{\partial \rho(p,T)}{\partial T}\left[-\frac{\dfrac{\partial E(p,T)}{\partial p} - \dfrac{p}{\rho^2(p,T)}\dfrac{\partial \rho(p,T)}{\partial p}}{\dfrac{\partial E(p,T)}{\partial T} - \dfrac{p}{\rho^2(p,T)}\dfrac{\partial \rho(p,T)}{\partial T}}\right] \qquad (8)$$

### III. EXPERIMENT

#### A. Measuring setup

For the measurements of the phase velocity of longitudinal ultrasonic waves the measuring setup was developed in the Institute of Fundamental Technological Research, Warsaw, Poland[13]. A special mounting of ultrasonic transducers in the high-pressure chamber was designed and fabricated to obtain a low level of parasitic ultrasonic signals. Longitudinal ultrasonic waves were excited and detected using 5 MHz LiNbO$_3$ (Y$_{36}$ cut) plates (Boston Piezo-Optics Inc. USA).

High-pressure chamber was designed and fabricated in the Institute of Physics at Warsaw University of Technology. High pressure was generated in a thick-walled cylinder with a simple piston and Bridgman II sealing system. The piston-cylinder assembly was working with a 20-ton hydraulic press, driven by a hand-operated pump. For pressure measurement 75 Ω manganin transducer was used. The piezoelectric transducers and manganin transducer were connected with the external measuring setup by an electrical multi-channel lead-through.

The sending $LiNbO_3$ transducer was driven by the TB-1000 pulser-receiver computer card (Matec, USA). The pulser generated the rf tone burst with a frequency 5 MHz. The longitudinal wave impulse generated by the sending transducer propagated in the investigated liquid and was detected by the receiving transducer. The PDA-1000 digitizer card (Signatec, USA) sampled and digitized the signals received by the transducer and amplified by the TB-1000 receiver. The stored signals were then analyzed by computer software. The time of flight of the ultrasonic pulses was evaluated by applying the cross-correlation method[14]. The cross-correlation method is a global differential method. The cross-correlation method does not depend on the trigger level and delays in cables and amplifiers.

**B. Experimental results**

In Fig. 1 phase velocity of longitudinal acoustic wave in triolein in function of hydrostatic pressure is presented. A triolein $(C_{17}H_{33}COO)C_3H_5$ is a triglyceride and unsaturated fat that is used as a model liquid in investigations of vegetable oils.

The pressure was generated in 10 MPa steps then kept constant about 2-5 min. That allowed to control whether the system was reaching thermodynamic equilibrium. The phase velocity was increasing monotonically with applied pressure. The measurements of the phase velocity were carried out from atmospheric pressure up to 450 MPa. Above this value of pressure, the first-order phase transition occurs in triolein[13]. Measurements were performed at the temperature $T = 300\,K$. Sound velocity measurement is based on a through-transmission method which make use of two transducers placed at a distance $d$ between them. The sound velocity is determined from the transit time $t$ of an ultrasonic pulse through the liquid and the corresponding acoustic path length $d$, ($c = d/t$). One may conclude that the overall accuracy of the sound velocity measurements is better than $0.1\%$.

# IV. APPROXIMATION OF EXPERIMENTAL SOUND SPEED CURVES BY MEANS OF ANALYTICAL FORMULAS

This paper has two purposes. The first aim is to report our experimental results relating to the speed of sound measurements in triolein in high pressure. The second aim is to search for such phenomenological expressions that give a good fit to our experimental results and, at the same time, depend on a small number of parameters.

For that purpose, the established by the authors Eq.8 for the sound speed in adiabatic condition has been used. That equation depend on the mass density and on the energy density per unit mass (both quantities are defined as functions of pressure and temperature) and is established here for the first time. The variables used in this equation are convenient for description of fluids under elevated pressure (a similar equation but formulated in terms of the mass density and the temperature has been developed in[9]).

There are (at least) two different ways of exploiting the Eq.8 for obtaining approximate analytic formula for $c^2(p)$. Both alternatives are compared in Appendix. In the first approach one looks for an „approximate" Gibbs function (depending on a small number of parameters) whereas in the second approach one tries to find individual corrections for $\rho(p,T)$ and $E(p,T)$. In our case, the second approach gives the better fit and therefore it has been applied in Eq.9 and Eq.10.

## A. Two-parameter approximation

The complete sound velocity data for the liquid can be fitted to a two-parameter analytical formula (see, Eq.A8 taken for $\gamma = 0$):

$$c^2 = \frac{[B + \lambda \cdot p] \cdot T}{\dfrac{B}{[B + \lambda \cdot p]} - \dfrac{[B - \lambda \cdot p]}{[B \cdot 3/2 + B + 2\lambda \cdot p]}} \qquad \text{(Eq.9)}$$

Figure 2 presents the fit of Eq.9 to the experimental data from Fig.1.
Pearson's correlation coefficient between the experimental curve from Fig.1 and that resulting from Eq.9 (2-parameter approximation) equals 0.99623.

**B. Three-parameter approximation**

Experimental data can be also fitted to a three-parameter analytical formula (see, Eq.A8):

$$c^2 = \frac{[B + \lambda \cdot p + 2\gamma \cdot p^2] \cdot T}{\frac{[B - 2\gamma \cdot p^2]}{[B + \lambda \cdot p + 2\gamma \cdot p^2]} - \frac{[B - \lambda \cdot p - 4\gamma \cdot p^2]}{[B \cdot 3/2 + B + 2\lambda \cdot p + 3\gamma \cdot p^2]}} \quad \text{(Eq.10)}$$

Figure 3 presents the fit of Eq10 to the experimental data from Fig.1. Correlation coefficient between the measured curve from Fig.1 and that resulting from Eq.10 (3-parameter approximation) equals 0.999651.

In order to compare different approximations, experimental data from Fig.1 have also been fitted to the first and second order polynomials, see Figs.4 and 5. Correlation coefficient between the measured curve from Fig.1 and its first order (2-parameter) polynomial approximation equals 0.98704, see Fig.4. Similarly, correlation coefficient between the measured curve and its second order (3-parameter) polynomial approximation equals 0.99882, see Fig.5. Fittings of the experimental curves to theoretical curves were made using the approximate procedures of the package Mathcad ®.

After comparing Figs.2-5 it can be seen that the best fit has been obtained by means of the 3-parameter approximation (Eq.10). Moreover, the two and three-parameter approximations (Eqs.9 and 10) give better fit than corresponding polynomial approximations with the same number of parameters. It is evident that, an optimal fit should be searched for each given number of parameters separately, because for more parameters it is easier to find a good fit, than for a smaller number of parameters.

**V. CONCLUSIONS**

A new formula (Eq.8) describing the velocity of acoustic waves in liquids in adiabatic conditions has been established. This formula displays the dependence of sound velocity on the mass density $\rho(p,T)$ and internal energy per unit mass $E(p,T)$ of liquid expressed in function of pressure $p$ and temperature $T$ as independent variables.

Velocity of longitudinal acoustic waves propagating in triolein was measured in the pressure range from atmospheric pressure to 450 MPa.

For experimental data, phenomenological approximate formulas (Eqs.9 and 10) have been proposed. These formulas have two versions (depending on the 2 and 3 parameters). Both versions have been obtained with the help of the new formula (Eq.8) for the sound speed in adiabatic conditions (pressure and temperature are taken as the independent variables). These 2 and 3-parameter approximations are more general and give a more accurate fit to experimental data than polynomial approximations with the same number of parameters.

It was found that the applied heuristic method that treats functions $\rho(p,T)$ and $E(p,T)$ independently allows a better fit for a small number of parameters than the method using strictly the Gibbs function that depends on the same number of parameters.

From an analysis of the experimental results of this investigation, it can be concluded that the established analytical formulas for sound speed $c^2$ can be useful for modeling the behavior of liquids under high pressure.

[13] P. Kiełczyński, M. Szalewski, A.J. Rostocki, M. Zduniak, R.M Siegoczyński, and A. Balcerzak, "Investigation of high-pressure phase transitions in vegetable oils by measuring phase velocity of longitudinal ultrasonic waves", 2009 IEEE Int. Ultrason. Symp. Proc., Rome 2009, 1563-1566.

[14] S. Sugasawa, "Time difference measurement of ultrasonic pulses using cross-correlation function between analytic signal", Jpn. J. Appl. Phys. **41**, 3299-3307 (2002).
12

**APPENDIX**

The Eq.8 for the sound speed in adiabatic conditions depends on the mass density $\rho(p,T)$ and on the internal energy density per unit mass $E(p,T)$. Both these quantities can be expressed in terms of the Gibbs function.

For an ideal gas, the explicit form of the Gibbs function is the following:

$$G_1(p,T,A,B) = (A+B)T - T[(A+B)\ln T - B \ln p] \qquad (A1)$$

where: $A$ and $B$ are parameters.

For our experiment, it is possible to search for such a form of the Gibbs function that gives an optimal fit to experimental results and, at the same time, depends on a small number of parameters. Our numerical experiments suggest that the best approximation for the Gibbs function is

$$G_2(p,T,A,B,\lambda,\gamma) = G_1(p,T) + \lambda \cdot p \cdot T + \gamma \cdot p^2 \cdot T \qquad (A2)$$

where $G_1(p,T)$ is the Gibbs function for an ideal gas and the additive correction depends on the two additional parameters $\lambda$ and $\gamma$.

For the Gibbs function (A2) it is possible to determine the mass density $\rho$ and the internal energy per unit mass $E$ according to general rules:

$$\rho = \frac{1}{\frac{\partial G_2}{\partial p}} = \frac{p}{T \cdot [B + \lambda \cdot p + 2\gamma \cdot p^2]} \qquad (A3)$$

$$E = G_2 - p \frac{\partial G_2}{\partial p} - T \frac{\partial G_2}{\partial T} = A \cdot T - \lambda \cdot p \cdot T - 2\gamma \cdot p^2 \cdot T \qquad (A4)$$

After inserting (A3) and (A4) into (Eq.8) one obtains:

$$c^2 = \frac{[B + \lambda p + 2\gamma p^2]T}{\dfrac{[B - 2\gamma p^2]}{[B + \lambda p + 2\gamma p^2]} - \dfrac{[B + \lambda p]}{[A + B + \gamma p^2]}} \qquad (A5)$$

Our numerical experiences have shown that one can not get a good fit of this equation to experimental data.

On the other hand, in order to find a good fit of the approximate expressions for the sound speed to experimental data, one can check an alternative (heuristic) procedure: it is possible to search for individual corrections for $\rho(p,T)$ and $E(p,T)$ and insert them directly into Eq.8. It can be checked that after taking:

$$\rho = \frac{p}{T \cdot [B + \lambda \cdot p + 2\gamma \cdot p^2]} \qquad (A6)$$

$$E = A \cdot T + \lambda \cdot p \cdot T + 2\gamma \cdot p^2 \cdot T \qquad (A7)$$

and after inserting the above expressions into (Eq.8) one obtains the following analytical formula

$$c^2 = \frac{[B + \lambda \cdot p + 2\gamma \cdot p^2] \cdot T}{\dfrac{[B - 2\gamma \cdot p^2]}{[B + \lambda \cdot p + 2\gamma \cdot p^2]} - \dfrac{[B - \lambda \cdot p - 4\gamma \cdot p^2]}{[A + B + 2\lambda \cdot p + 3\gamma \cdot p^2]}} \qquad (A8)$$

that gives a good fit to experimental data (see, Chapter 4).

In order to reduce the number of parameters, we can assume that: $A = 3/2 \cdot B$. Consequently, Eqs.9 and 10 have been derived from Eq.A8 using this relation, i.e., $A = 3/2 \cdot B$.

In addition to the Gibbs function expressed by Eq.A2 also some other forms of Gibbs functions (depending on a small number of parameters) have been applied (e.g., $G = G_1 + \lambda \cdot p/T$) in the procedure of approximation of experimental data by approximate analytical relations. Although these forms of the Gibbs function lead to analytical formulas that approximate the experimental data satisfactorily, however using the approximate expressions resulting from the heuristic method, i.e., equation A8, we get much better approximations to the experimental data.

Figure captions

Fig.1. Plot of the measured speed of sound in triolein as a function of hydrostatic pressure, $T = 300\,K$.

Fig.2. 2-parameter approximation (Eq.9) of the measured curve of speed of sound in triolein versus hydrostatic pressure, $T = 300\,K$.

Fig.3. 3-parameter approximation (Eq.10) of the measured curve of speed of sound in triolein as a function of hydrostatic pressure, $T = 300\,K$.

Fig.4. First-order polynomial approximation (2 parameters) of the measured curve of speed of sound in triolein as a function of hydrostatic pressure $(c = a_0 + a_1 p)$.

Fig.5. Second-order polynomial approximation (3 parameters) of the measured curve of speed of sound in triolein as a function of hydrostatic pressure $(c = a_0 + a_1 p + a_2 p^2)$.

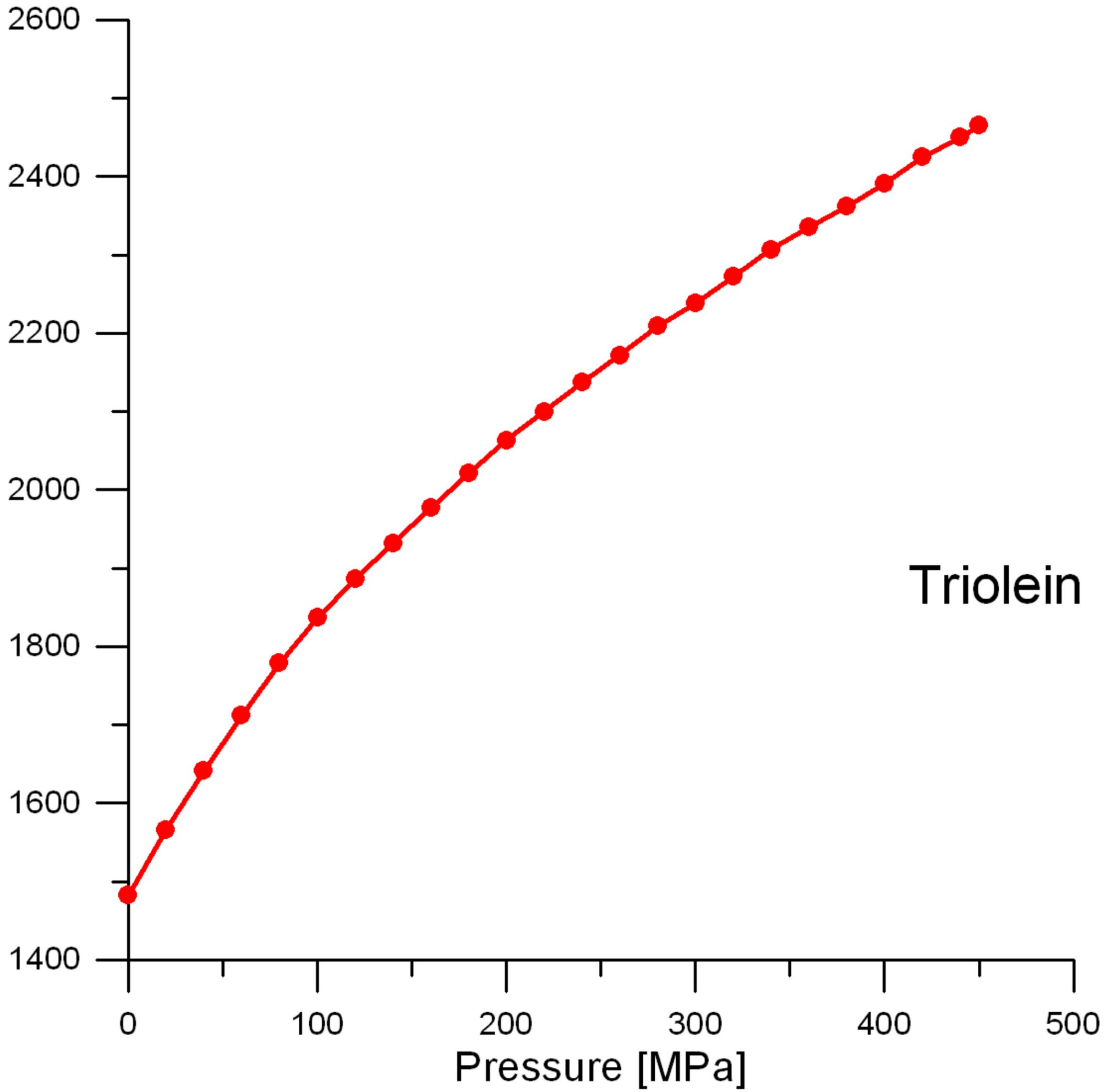

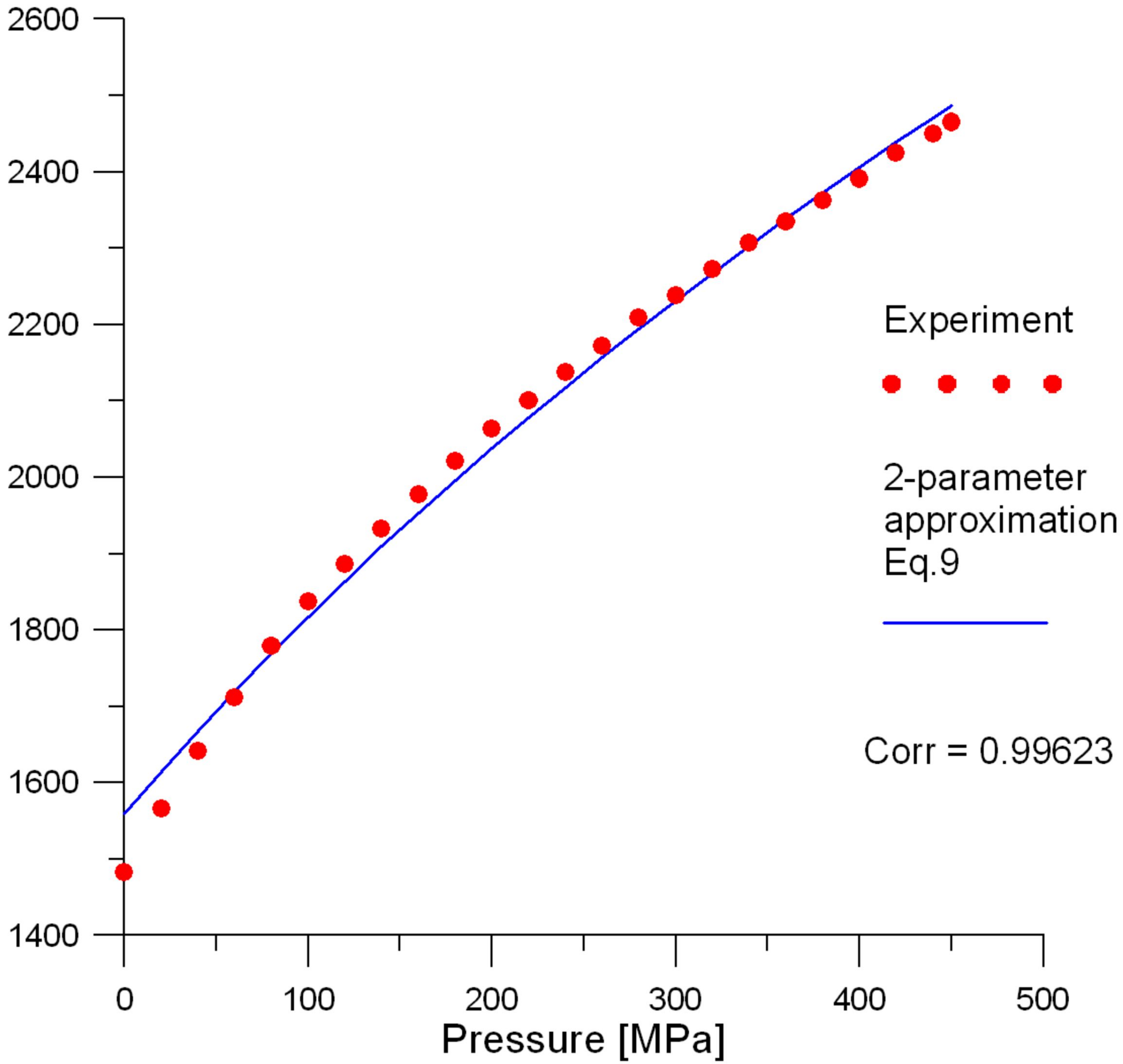

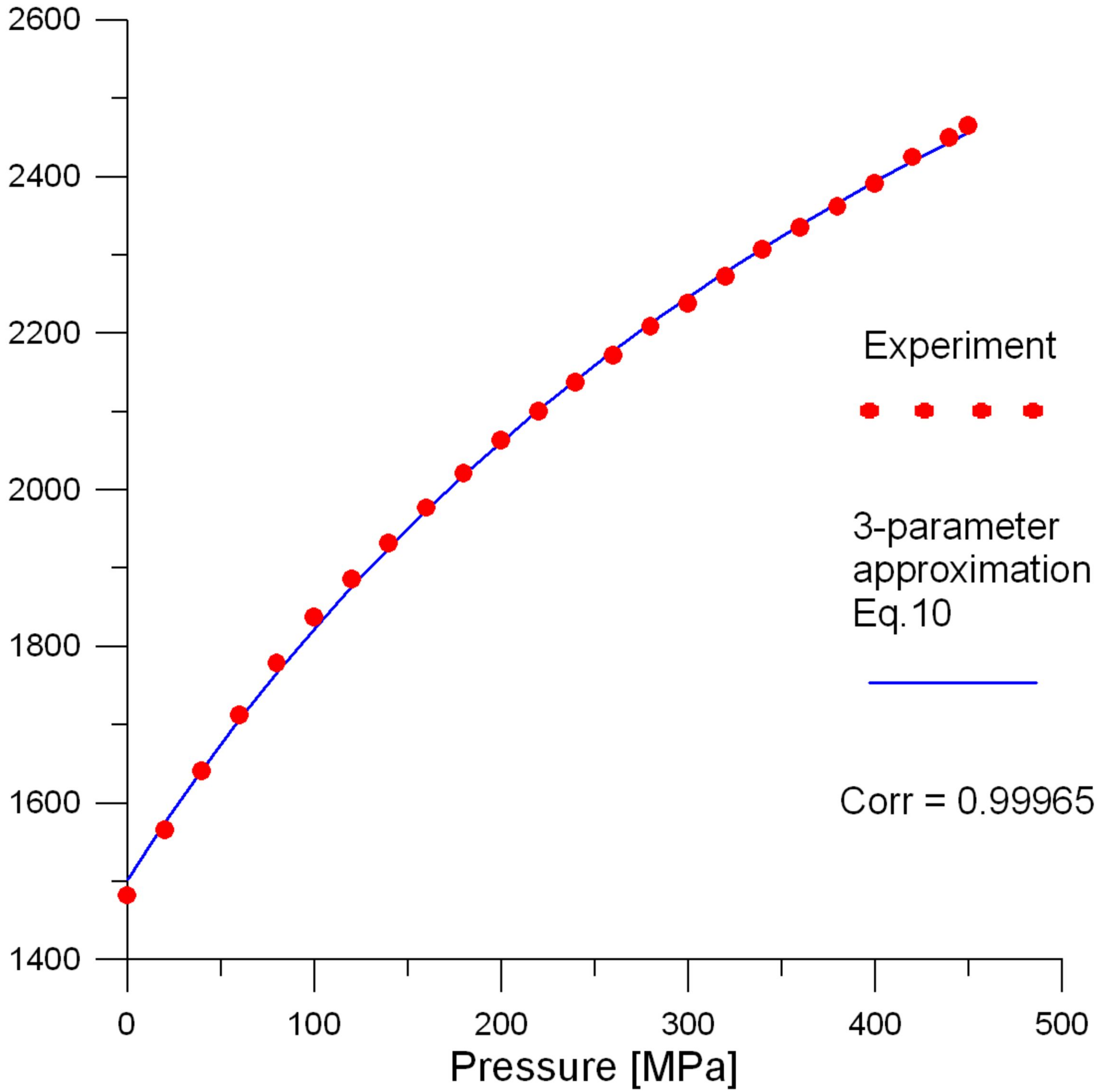

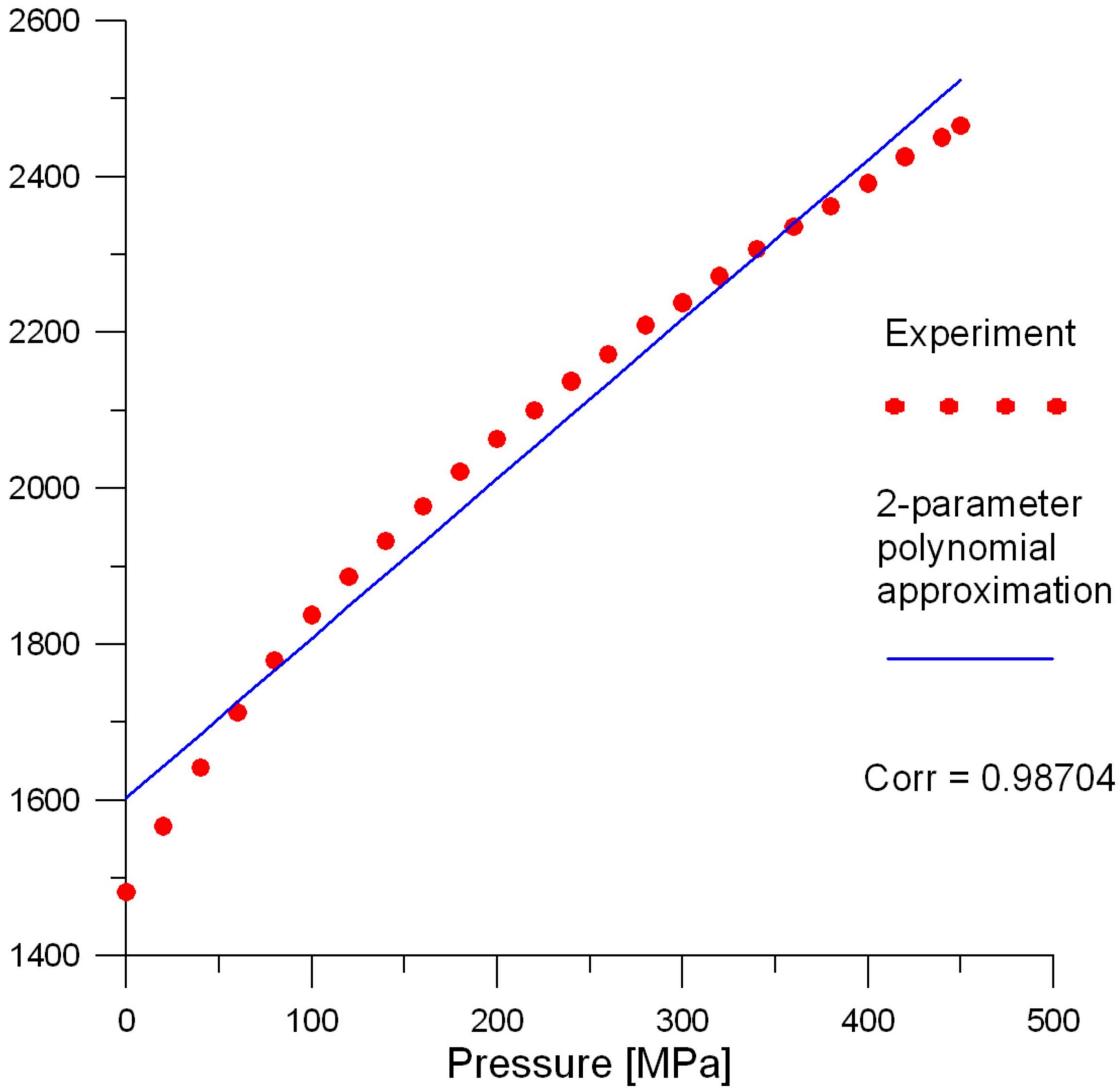

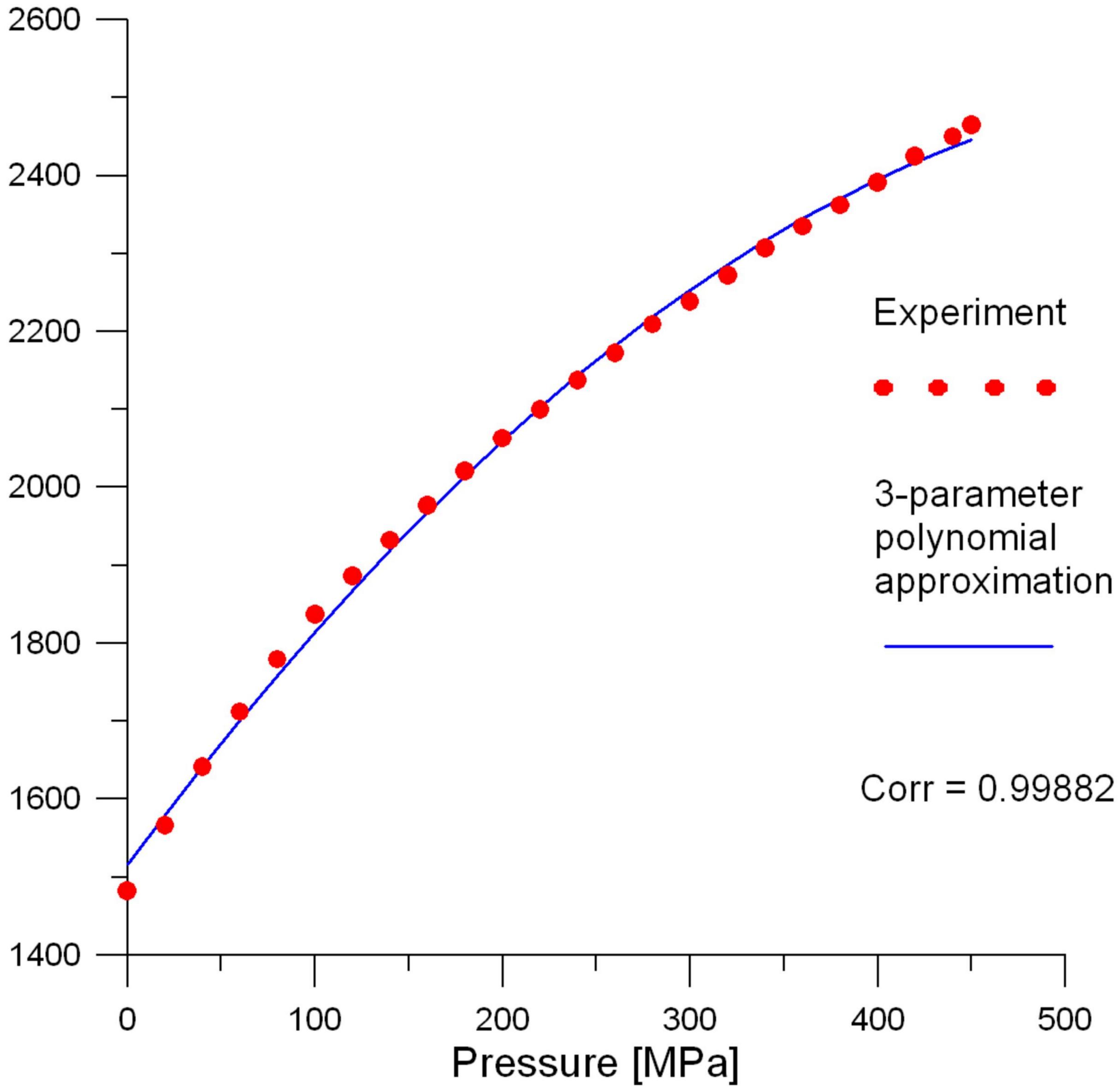